\documentstyle[preprint,aps,psfig]{revtex}
\tightenlines

\def\beq{\begin{equation}}
\def\eeq{\end{equation}}
\def\bea{\begin{eqnarray}}
\def\eea{\end{eqnarray}}
\def\nnu{\nonumber}

\def\al{\alpha}
\def\be{\beta}
\def\gam{\gamma}
\def\dta{\delta}
\def\eps{\epsilon}
\def\tta{\theta}

\def\om{\omega}

\def\Dta{\Delta}
\def\Gam{\Gamma}
\def\Om{\Omega}

\def\ham{{\cal H}}
\def\ket#1{|#1\rangle}
\def\bra#1{\langle#1|}

\def\avg#1{\langle#1\rangle}
\def\mel#1#2#3{\langle#1|#2|#3\rangle}

\def\intzi{\int_0^\infty}
\def\dint{\int\!\!\!\int}

\def\bc{{\bf c}}

\def\bz{{\bf z}}
\def\bq{{\bf q}}

\def\Mn12{Mn$_{12}$}
\def\Fe8{Fe$_8$}
\def\ep#1#2{\eps_{#1#2}}
\def\hsp{\ham_{\rm sp}}
\def\gab{\gam_{\rm abs}}

\def\tD{\widetilde D}
\def\ta12{{\tilde a}_{12}}
\def\gdir{\gam_{\rm dir}}
\def\gi#1{\gam_{{\rm in},#1}}
\def\gin{\gam_{\rm in}}

\begin{document}
\draft

\title{Lattice Distortion Mediated Paramagnetic Relaxation in High-Spin
High-Symmetry Molecular Magnets}

\author{Anupam Garg}
\address{Department of Physics and Astronomy, Northwestern University,
Evanston, Illinois 60208}

\date{May 11, 1998}

\maketitle

\begin{abstract}
Field-dependent maxima in the relaxation rate of the magnetic molecules \Mn12-Ac
and \Fe8-tacn have commonly been ascribed to some
resonant tunneling phenomenon. We argue instead that the
relaxation can be understood as purely due to phonons. The rate maxima arise because of a
Jahn-Teller-like distortion caused by the coupling of lattice phonons to degenerate
Zeeman levels of the molecule at the top of the barrier. The binding energy of the
distorted intermediate states lowers the barrier height and increases the relaxation
rate. A nonperturbative calculation of this effect is carried out for a model system.
An approximate result for the field variation near a maximum is found to agree
reasonably with experiment.

\end{abstract}

% \widetext

In the last five years, a remarkable and novel type of paramagnetic relaxation has
been observed in certain molecular crystals \cite{ns,fs,tlbd,sop}. The best known,
a \Mn12-acetate complex (or
just \Mn12 for short), has a ground state spin $S=10$, and a four-fold symmetry
axis $\bz$, which coincides with the $\bc$ axis of the tetragonal crystal that it forms.
It is found that if the spin is first polarized along $\bc$, and then allowed
to relax by reversing the field to a value $H$, the relaxation rate is a non monotonic
function of $H$, showing maxima at roughly equally spaced fields $H_n= nH^*$, with
$H^* = 0.44$ T. Hysteresis loops measured with a slow field sweep show
sharp jumps in the magnetization $M$ at the same reverse fields $H_n$.
The relaxation rate $\Gam$ obeys an Arrhenius law with a barrier of $\sim62$ K when $H=0$. 
Similar effects have been seen in
[(tacn)$_6$Fe$_8$O$_2$(OH)$_12$]$^8+$, though not as sharply.

In this Letter we contend that at present there is not even a qualitative understanding
of these rate maxima, and attempt to fill this gap. We begin by
reviewing the basic background theory, which is due to Villain, Hartmann-Boutron, Sessoli,
and Rettori (VHSR) \cite{vhsr}.
The principle terms in the spin Hamiltonian are
\beq
\ham_s = -DS_z^2 - hS_z, \label{hams}
\eeq
where $h = g\mu_B H$. The next order terms, $S_z^4$ and $S_{\pm}^4$, are small and
have been ignored.
The spin states $\ket m$, where $S_z\ket m = m \ket m$, have energies
$\eps_m = -Dm^2-hm$. Suppose the initial spin state is $\ket{-S}$, and
$\eps_{-S} > \eps_S$.  As explained by VHSR, the spin relaxes via a multistep
Orbach process. It absorbs phonons and climbs from $m=-S$
to $-S+1$, $-S+2$, $\ldots$, until it reaches the highest level $m^*$, and then descends
to levels $m^* +1$, $m^*+2$, $\ldots$, $S$ by emitting phonons. The 
occupation probabilities $\{p_m\}$ obey a set of master equations
\beq
{dp_m \over dt} = -\sum_n\gam_m^n p_m + \sum_n\gam^m_n p_n, \label{rteqn}
\eeq
where $\gam_m^n$ is the rate or probability per unit time for the $m \to n$ transition.
To calculate $\{\gam_m^n\}$, VHSR use Fermi's
golden rule with a standard spin-phonon interaction
\beq
\ham_{\rm sp} = \sum_\al g_{ijkl}\,\om_{\al}^{-1/2}
                e_{\al i}q_{\al j}(a_{\al} + a^{\dag}_{\al})(S_kS_l + S_lS_k),
    \label{hsp}
\eeq
where $\al$ labels the phonon modes with creation and anihilation operators
$a^{\dag}_\al$ and $a_\al$, ${{\bq}}_\al$, ${{\bf e}}_\al$, and $\om_\al$
 are the wavevector, polarization, and frequency, and $i$, $j$, etc. are
Cartesian indices.  The tensor $g_{ijkl}$ is only weakly
$\bq$ dependent. With $\ep mn \equiv \eps_m - \eps_n$, we obtain
\beq
\gam_m^n = K |V_{mn}|^2 \times \cases{
                               \ep nm^3 n(\ep nm), & $\eps_m < \eps_n$, \cr
                               \ep mn^3 [1+ n(\ep mn)], & $\eps_m > \eps_n$. \cr} 
\label{fgr}
\eeq
Here, $K$ is a constant dependent on the sound speed, the density of the solid, and the
couplings $g_{ijkl}$, $V_{mn}$ denotes a spin matrix element such as
$\bra m S_kS_l \ket n$, and $n(\eps)$ is the Bose function
$n(\eps) = (e^{\be\eps} - 1)^{-1}$,
with $\be = 1/k_B T$. VHSR solve these rate equations and find that for $H=0$,
\beq
\Gam = K |V_{10}|^2 D^3 \exp(-\be DS^2)/(1-e^{-\be D}). \label{Vrt}
\eeq
The factor $e^{-\be DS^2}$ yields the expected activated behavior with an energy barrier
$DS^2$. The spin phonon interaction required to fit the observed rate is somewhat large,
but not unreasonable.

Since the relaxation is a sequential process in this picture, $\Gam$
is limited by the weakest link in the chain \cite{fna}. This leads to a
disaster whenever the field is such that the topmost two levels, $m$ and $m'$, say,
are degenerate. For then, by Eq.~(\ref{fgr}), $\gam_{m'}^m$ and $\gam_m^{m'}$
vanish as $k_B T \eps_{mm'}^2$. One link in the chain is broken, and so
$\Gam$ must vanish completely!

In fact this conclusion is not correct in detail. To see this, we first note
that the $S_xS_z$ and $S_yS_z$ terms in Eq.~(\ref{hsp})
connect levels with $\Dta m =1$, while $S_xS_y$ like terms allow
$\Dta m =2$ processes. (The $\Dta m =0$ terms lead to ignorable level shifts.)
Second, there are two types of level crossings: $m$ and $m+1$ coincide
($\Dta m =1$ or ``odd" crossings) when
$h=h_{-(2m+1)}$, and $m$ and $m+2$ coincide ($\Dta m = 2$ or ``even" crossings)
when $h = h_{-(2m+2)}$, where $h_j\equiv jD$. Since the $\Dta m= 2$ channel is open
at a $\Dta m = 1$ crossing, and vice versa, relaxation is still possible.
Nevertheless, $\Gam$ should be a {\it minimum}
whenever $h=h_n$ for any $n$, even or odd, as one channel is closed off. This
still poses a problem, as these are just the field values where the observed
$\Gam$ is a {\it maximum}. Taking $D = 0.57$ K from electron
spin resonance (ESR) experiments \cite{bgs},
we get $H_j  = 0.42j$ T, which agrees quite well with
the measured jump fields in the hysteresis loops.

Given the above agreement, it is extremely tempting to argue that there is a resonant
coupling between degenerate energy levels near the top of the barrier \cite{fnb}
due to some perturbation, such as hyperfine or dipole-dipole interaction, or terms
such as $S_+^4 + S_-^4$ in the spin Hamiltonian \cite{many,hpv}.
By and large, all these mechanisms are either too weak
to explain the observed widths of 0.03--0.05 T in the relaxation rate maxima,
and/or run into difficulties with selection rules. The hyperfine field,
e.g., has an estimated rms strength of ~0.015 T, and must act in second
order to connect levels with $\Dta m =2$, which implies a width $\sim10^{-3}$ T.
Similarly, the dipolar field is estimated as $\sim0.01$ T. Not only is this
again a little small, but an 
experiment on a frozen solution of \Mn12 \cite{rs} with greater intermolecular
separation than the crystal shows {\it faster} relaxation, exactly the
opposite of what would be expected. The $S_{\pm}^4$ terms cannot explain the\
$\Dta m =1$ maxima. The strongest evidence that a perturbation involving
only the spin is not responsible for the rate maxima comes from
experiments in which the field is applied at an angle $\tta$
to the $z$ axis \cite{fs2,ltbb}. Since the perturbation $-h_x S_x \propto \tta$
for small $\tta$, $\Gam$ should rise dramatically as $\tta$ increasess from
say $0.5^\circ$ to $10^\circ$ \cite{fnc}, with $H_z$ held fixed at $H_n$. However, as can be
seen from Fig.~4 of Ref.~\cite{ltbb}, the jump $\Dta M$ hardly changes
at all for $H_z=H_1$, and increases only by 2 at $H_z=H_2$ as $\tta$ is varied from 0 to
$10^\circ$. Further, the width of the jump at $H_1$ is essentially unchanged even as
$\tta$ is increased to $44^\circ$, whereas
a simple anticrossing would yield a width varying as $D\sin\tta$.

How, then, does the magnetization relax when $H=H_n$?  In our view, the causative
agency is still the spin-phonon interaction but it now plays a dual role. Specifically, the
part of $\hsp$ that couples two levels can not be treated perturbatively when these two
levels are degenerate. One way to see this is to map the degenerate levels onto the
$s_z = \pm 1/2$ levels of a pseudospin with $s=1/2$, to write $\hsp$ and $\ham_p$ as
$\sum c_\al s_x x_\al$ and $\sum\om_\al(p_\al^2 + x_\al^2)/2$, where $x_\al$ and
$p_\al$ are oscillator position and momentum coordinates for mode $\al$. It is then
obvious that the lowest energy eigenstates are those of $s_x$, not $s_z$, and that in
these states the oscillators have a non zero mean displacement,
$\avg{x_\al} =-c_\al  s_x/\om_\al$.  The binding energy $\Om$ due to this displacement,
effectively leads to a reduction of the net barrier, and hence
to an enhanced relaxation rate.
\begin{figure}[b]
\centerline{\psfig{figure=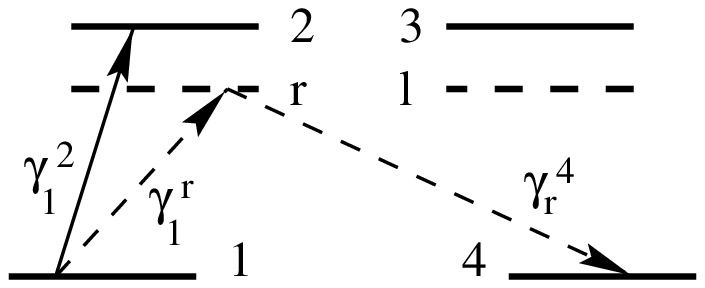}}
\caption{\\ Upper four magnetic levels in the degenerate case, without (solid lines) and
with (dashed lines) lattice distortion.}
\end{figure}
In the rest of the Letter we shall sketch a calculation of this effect for a model problem,
in which the spin-phonon interaction is not the most general one possible, but which would
also yield a vanishing rate at degeneracy if treated perturbatively. We will show that
phonons alone give a nonzero rate, which in fact exceeds that which
would be guessed from a Arrhenius law which the barrier varies smoothly with
$H$ and the prefactor is taken as constant.  We will also show that the rate is a local
maximum at degeneracy, with an approximate field dependence that describes the data
on \Mn12 fairly well.

Our model focuses on the four energy levels near the top of the barrier (see
Fig. 1), which we label 1, $\ldots$, 4.
The full Hamiltonian $\ham$ is the sum of spin,
spin-phonon, and phonon terms:
\bea
\ham_s &=&  \sum_{j=1}^4 \eps_j \ket j \bra j, \label{hs2} \\
\hsp   &=&  \sum_{\al, j} a_{j,j+1}(t_{j,j+1} + t_{j+1,j}) c_\al x_\al, \label{hsp2} \\
\ham_p &=&  \sum_{\al} {\om_\al \over 2} (p_\al^2 + x_\al^2), \label{hp}
\eea
where $t_{ij} = \ket i \bra j$, and we have set $\hbar =1$. We will take
$\eps_1 \approx \eps_4$, $\eps_2 \approx \eps_3$, and make the inessential
simplification $a_{34} = a_{12}$. Also, we choose a spectral density
\beq
J(\om) \equiv {\pi\over 2} \sum_\al c_\al^2 \dta(\om-\om_\al)
       = B \om^3 \exp(-\om/\om_c), \label{Jom}
\eeq
which describes a coupling to accoustic phonons as per by Eq.~(\ref{hsp}).
The cutoff $\om_c$ is like a Debye frequency. Defining $D = \ep 21$, we assume
$\om_c \gg D,\ k_BT$,
and also that the maximal lattice distortion binding energy $\Om$ (realized
in the case of exact degeneracy $\eps_2 = \eps_3$)
obeys $\Om \ll D$. This energy is given by
\beq
\Om  =  {a_{23}^2 \over \pi} \int_0^\infty d\om {J(\om) \over \om} = a_{23}^2 \sum_\al {c_\al^2 \over 2 \om_\al}. \label{solv}
\eeq
Note that since all oscillators contribute to $\Om$ and not just those near $\om=0$,
the high frequency details of $J(\om)$ can significantly affect $\Om$.

We now wish to calculate $\gam_1^4$, the net rate at which the spin, initially in
state $\ket 1$, relaxes to state $\ket 4$, at a temperature $T$. 
For future use and for comparison with Eq.~(\ref{fgr}), we record here the result of a
Fermi golden rule calculation of the $1 \to 2$ transition rate:
\beq
\gab^0 = 2 a_{12}^2 J(D) n(D) \approx 2 B a_{12}^2 D^3/(e^{\be D} -1).
   \label{rt0}
\eeq
A similar golden rule calculation gives $\gam_2^3 = 0$, and as
discussed earlier, utterly fails to explain how the spin relaxes. The
correct procedure is to first find the exact eigenstates of the
{\it full} Hamiltonian in the 23 subspace,
$\ham_{23} = P_{23}\ham P_{23}$, $P_{23} = \ket 2 \bra 2 + \ket 3 \bra 3$,
and then find the absorption and emission rates to and from these intermediate
states by the golden rule.
The absorption rate from state 1 so calculated can be formally written as
\beq
\gab = 2\pi \sum_\mu p_\mu
          \mel{1,\mu}{V^{\dag} \dta(\eps_1 +\eps_\mu - \ham_{23}) V}{1,\mu},
      \label{gab}
\eeq
where $V = a_{12} \sum_\al c_\al x_\al \ket 2 \bra 1$, and $\mu$ labels the initial phonon
states, with energy $\eps_\mu$, and occurrence probability
$p_\mu \propto \exp(-\be\eps_\mu)$. The emission rate can be similarly expressed.

Formula (\ref{gab}) can be greatly simplified when $\eps_2 = \eps_3$.
Fourier transforming the energy $\dta$-function, and expressing the sum
over $\mu$ in terms of the density matrix $\exp(-\be\ham_p)$, we obtain a trace over phonons
which involves only Gaussian integrals. A lengthy but simple analysis yields
\beq
\gab = a_{12}^2 \int_{-\infty}^{\infty} \!\! d\tau
       e^{-i[D \tau - a_{23}^2 R(\tau)]} 
        \left[ S_1(\tau) + a_{23}^2 S_2^2(\tau) \right],
   \label{rtex}
\eeq
where
\bea
S_1(\tau) &=& {1 \over \pi} \intzi \!\! d\om \,J(\om) (f_a + f_e),  \label{S1} \\
S_2(\tau) &=& {1 \over \pi} \intzi \!\! d\om {J(\om) \over \om}
               (1 + f_a - f_e), \label{S2} \\
R(\tau)  &=& {1 \over \pi} \intzi \!\! d\om {J(\om) \over \om^2}
               \left[\om\tau + i\coth{\be \om \over 2}
                        - i(f_a + f_e) \right], \label{Rt}
\eea
with $f_a = e^{i\om\tau}n(\om)$, and $f_e = e^{-i\om\tau}[1+ n(\om)]$.

The essential aspects of the integral (\ref{rtex}) follow from
the behavior of $a_{23}^2 R(\tau)$ as $\tau \to \infty$.
The first term in Eq.~(\ref{Rt})
grows linearly with $\tau$, the second tends to a constant, and the last two vanish
as $\tau \to \infty$. We make this explicit by writing
\beq
R(\tau) = a_{23}^{-2} (\Om\tau + iW) -i R'(\tau), \label{Rt2}
\eeq
where $\Om$ is the binding energy (\ref{solv}),
\beq
W = {a_{23}^2 \over \pi} \int_0^\infty d\om {J(\om) \over \om^2} 
          \coth{\be\om \over 2}, \label{DW}.
\eeq
and $R'(\tau) = \int d\om J(\om)(f_a+f_e)/\pi\om^2$.
The term $\Om\tau$ clearly must be
combined with $D\tau$ in the exponential factor in Eq.~(\ref{rtex}).
This point makes it obvious that the relevant transition energy when $\ep32 = 0$
is not the bare $D$, but rather the reduced quantity
\beq
\tD = D - \Om. \label{tD}
\eeq
Likewise, the term $W$ reduces the coupling constant $a_{12}$ to
$\ta12 = a_{12}\exp(-W/2)$. This is nothing but the
Franck-Condon overlap factor between the unpolarized and polarized lattice states
accompanying the spin states 1 and 2 (or 3), respectively.

The explicit form (\ref{Jom}) for $J(\om)$ yields $\Om = 2Ba_{23}^2 \om_c^3 /\pi$, and
$W \approx Ba_{23}^2\om_c^2/\pi = 2\Om/\om_c$ for $k_BT \ll \om_c$. Since
$\Om < D$ and $D \ll \om_c$ by assumption, $W \ll 1$.

We can further approximate Eq.~(\ref{rtex}) by expanding the remaining factor
$\exp[a_{23}^2 R'(\tau)]$ in powers of $a_{23}$ up to $O(a_{23}^2)$, which
is justified as long as
$a_{23}$ is not too large, and reflects what we mean by the ``weak" non-perturbative
limit. If this is done, $\gab$ can be written as the sum of
``direct" and ``indirect" parts \cite{fnd}, 
\bea
\gdir &=& \ta12^2 \int_{-\infty}^{\infty} d\tau e^{-i\tD \tau} S_1(\tau)
      = 2\ta12^2 J(\tD) n(\tD),   \label{gdir}  \\
\gin &=& \ta12^2 a_{23}^2\int_{-\infty}^{\infty} d\tau
       e^{-i\tD \tau} \left[ S_2^2(\tau) + S_1(\tau) R'(\tau) \right]. \label{gin}
\eea
The last result for $\gdir$ follows by first doing the $\tau$ integral,
and noting that $\tD > 0$, so only the absorption term $f_a$ in 
Eq.~(\ref{S1}) contributes. A similar procedure shows that $\gin$ can be expressed
as the sum of three parts,
\bea
\gi1 &=& 4 \ta12^2 {\Om\over \tD} J(\tD) n(\tD), \label{gin1} \\
\gi2 &=& {1\over \pi} \ta12^2 a_{23}^2 \tD^2
             \dint \!d\om_1 d\om_2 {J(\om_1) J(\om_2) \over \om_1^2 \om_2^2} 
             n(\om_1) n(\om_2) \dta(\om_1 + \om_2 - \tD), \label{gin2} \\
\gi3 &=& {2\over \pi} \ta12^2 a_{23}^2 \tD^2
             \dint \!d\om_1 d\om_2 {J(\om_1) J(\om_2) \over \om_1^2 \om_2^2} 
             n(\om_1) [1+n(\om_2)] \dta(\om_1 - \om_2 - \tD). \label{gin3}
\eea
It can now be shown that $\gi2/\gi1 \sim (\tD/\om_c)^3$ for $k_BT \ll \tD$, and
$\sim  \tD^2 k_BT/\om_c^3$ for $k_BT \gg \tD$,
while $\gi3/\gi1 \sim \tD (k_BT)^2/\om_c^3$ for all $T$. It is assumed that
$k_BT \ll \om_c$ always. Dropping $\gi2$ and $\gi3$ completely, we
obtain the compact approximation [compare with Eq.~(\ref{rt0})], 
\bea
\gab &\approx& 2\ta12^2 (1+ 2\Om/\tD) J(\tD) n(\tD)  \nnu \\
     &=& 2B\ta12^2 \tD^3 (1+2\Om/\tD)/(e^{\be\tD} -1). \label{gab2}
\eea
Note that $\gin \ll \gdir$ because $\Om \ll D$, which
renders the expansion in $a_{23}$ sensible.
The condition $\Om \ll D$ is actually basic to the very formulation of
our calculation, for otherwise
the levels $\ket 1$ and $\ket 4$ will also induce significant polarization of
the lattice, and treating them as bare levels is unjustified.

The last step is to find the over all rate $\gam_1^4$.
Let us do this first for the symmetric case, where $\ep 32 = \ep 41 =0$.
Then, phonon absorption starting from state $\ket 1$ takes the spin to states
$\ket{r,l} \equiv 2^{-1/2}(\ket 2 \pm \ket 3)$. The individual
rates for these transitions clearly obey $\gam_1^r = \gam_1^l = \gab/2$.
Liekwise the rates for phonon emission from these levels to level 4 are equal,
and given by $\gam_r^4 = \gam_l^4 = e^{\be\tD}\gab/2$ using detailed balance.
Following Ref.~\cite{vhsr}, one finds
\beq
\gam_1^4 = \gab. \label{g14}
\eeq

We thus see that even with degenerate top levels, phonons can lead to spin
relaxation. In fact the rate (\ref{g14}) exceeds
that given by a classical Arrhenius law in which the field dependence is smooth
and enters only through the overall energy barrier. In making this comparison, let us
also incorporate the lower $2S -3$ energy levels into the picture. The net rate is
again determined by the slowest step, and at low temperatures, $\exp(-\be\tD) \ll 1$,
it can be found by
multiplying $\gam_1^4$ by the Boltzmann factor for level 1. Ignoring factors of order
unity, and writing the matrix element as $|V|^2$, we have
\beq
\Gam \simeq K|V|^2 \tD^2(\tD + 2\Om) e^{-[W + \be(U-\Om)]}, \label{Myrt}
\eeq
where $U = \eps_1 - \eps_{-S}$ is the full energy barrier.
This should be compared with a ``naive" rate, $\Gam^0 \simeq K|V|^2 D^3 e^{-\be U}$.
It is not hard to see that $\Gam > \Gam^0$ provided
$k_BT < {\rm min}(\om_c, D^2/\Om)$ \cite{fne}.

We can also ask for the rate in the asymmetric case, when $\ep 32 \not= 0$.
We do not now have closed form answers, and our results are more approximate.
Let $\eps_2 = -\eps_3 = \zeta/2$, and
$\eps_1 = -\eps_4 = \zeta$, where $\zeta \ll D$. Lattice
polarization will now change $\zeta$ to $\zeta e^{-W}$, and the
spin states will be slightly rotated from $\ket {r,l}$. The main effect, however,
will be that the binding energy itself is reduced: $\Om \to \Om - c\zeta^2/\Om$,
where $c = O(1)$. This means that the center of gravity of the levels 2 and 3 will
not be lowered as much as in the degenerate case. Secondly, we can safely approximate
the rates $\gam_1^r$, etc. by the form (\ref{gab2}) (times 1/2), but with the correct
energy differences in the Bose factors. The rate $\gam_1^4$ can now be written as
a sum of rates through the parallel channels $\ket r$ and $\ket l$. A straightforward
analysis shows that the terms linear in $\zeta$ (due to differences in $\gam_1^r$
and $\gam_1^l$, e.g.) cancel. The chief surviving term is from the systematic reduction in
binding energy, leading to a factor $\exp(-c\be\zeta^2/\Om)$ in $\gam_1^4$. This
shows that the relaxation rate has a local maximum near the coincidence fields \cite{fnf}.

We conclude by asking how far our model applies to real \Mn12. The obvious omissions
are (a) $\Dta m =2$ processes, and (b) $t_{j,k} - t_{k,j}$ couplings in
Eq.~(\ref{hsp2}). Neither of these is expected to change the qualitative picture
given here. The key unknowns are the binding energy $\Om$, and whether $\Om \ll D$.
A look at Fig. 2 of Ref.~\cite{ltbb} shows that a field variation of the form 
$\Gam(H)\sim \exp(-c\be\zeta^2/\Om)$ with $\zeta = \Dta m(h-h_n)$, describes the rate
maxima quite well, and we find $\Om/c = $ 9, 6, 10, and 4 mK for $n =0$, 1, 2, and 3,
respectively. Direct estimates based, e.g., on Eq.~(B.3) of Ref.~\cite{hpv}, yield
$\Om$ values 0.5-20 mK, which are quite consistent given the uncertainties in $\om_c$ etc.
Direct experimental probes of $\hsp$ are clearly very important to testing our theory.
Suggestions for how this may be done, and more detailed calculations will be published
elsewhere.

I am indebted to B. Barbara, A. J. Leggett, C. Paulsen, M. Randeria, and R. Sessoli
for many useful discussions.
This work was done in part at the Institute for Nuclear Theory, University of
Washington, Seattle, and I thank the memebrs and staff of the Institute for their
warm hospitality. This work is supported by the National Science Foundation through
grant number DMR-9616749.


\begin{references}
\bibitem{fna} Thus, Eq.~(\ref{Vrt}) can be understood
as the Boltzmann factor for occupation of the level $m =-1$, times the
rate $\gam_{-1}^0$.
\bibitem{fnb} The simultaneous degeneracy of lower pairs of levels when $h=nD$
implied by Eq.~(\ref{hams}) is surely artificial. ESR indicates
a correction term $B(S_x^2 - S_y^2)$ for \Fe8 with $B = 0.046$ K, and $D'S_z^4$
for \Mn12 with $D' = -1.1\times 10^{-3}$ K. These lead to nondegeneracies for
the lower levels that are comparable or bigger than the perturbations used to couple
these levels. For \Mn12, e.g., we obtain $E_1 - E_{-2} = 12D'$ when $h=h_1$,
and $E_1 - E_{-3} = 48D'$ when $h=h_2$.
\bibitem{fnc} This argument would fail if the relaxation was already saturated at
$0.5^\circ$ (A.~J. Leggett, comment at seminar). This is unlikely to be so, because the
jump at $H_1$ is very small, and less than 3\% of the total $\Dta M$ for all
$\tta \le 10^\circ$.
\bibitem{fnd} The names are based on the prefactors in
$a_{j,j+1}$, and should not be interpreted too closely.
\bibitem{fne} It is quite likely that the ``naive" rate $\Gam^0(H)$ can not
quantitatively describe \Mn12 for {\it any} value of $H$ at low temperatures. The
comparison made here serves only to prove the point that lattice distortion effects
are large.
\bibitem{fnf} At {\it very} low temperatures, the systematic increase in $\gam_1^4$
from the decrease in $\ep 21$ may nullify this effect, but we do not have any
hard calculations to support this point.
\end{references}
\end{document}